\documentclass[journal]{IEEEtran}
\usepackage{graphicx,color}
\usepackage{cite}
\usepackage{setspace} 
\usepackage{amsmath}
\usepackage{amsmath,amsthm,amssymb}
\usepackage{multirow}
\usepackage{hhline}
\usepackage{epsfig}
\usepackage{subcaption}
\usepackage{epstopdf}
\usepackage{verbatim}
\usepackage{algorithm}
\usepackage{algorithmicx}
\usepackage{algpseudocode}
\usepackage{etoolbox}
\usepackage{cases}
\usepackage{csquotes}
\usepackage{enumitem}
\usepackage{mathtools,stmaryrd}
\SetSymbolFont{stmry}{bold}{U}{stmry}{m}{n}
\usepackage{bbm}
\usepackage{lettrine}
\usepackage{nicematrix}

\newcommand{\suchthat}{\;\ifnum\currentgrouptype=16 \middle\fi|\;}

\makeatletter
\newcommand*{\indep}{%
  \mathbin{%
    \mathpalette{\@indep}{}%
  }%
}
\newcommand*{\nindep}{%
  \mathbin{
    \mathpalette{\@indep}{\not}
  }%
}
\newcommand*{\@indep}[2]{%
  \sbox0{$#1\perp\m@th$}
  \sbox2{$#1=$}
  \sbox4{$#1\vcenter{}$}
  \rlap{\copy0}
  \dimen@=\dimexpr\ht2-\ht4-.2pt\relax
  \kern\dimen@
  {#2}%
  \kern\dimen@
  \copy0 
} 
\makeatother

\makeatletter
\newcommand*{\algrule}[1][\algorithmicindent]{%
  \makebox[#1][l]{%
    \hspace*{.2em}
    \vrule height .75\baselineskip depth .25\baselineskip
  }
}

\newcount\ALG@printindent@tempcnta
\def\ALG@printindent{%
    \ifnum \theALG@nested>0
    \ifx\ALG@text\ALG@x@notext
    \else
    \unskip
    \ALG@printindent@tempcnta=1
    \loop
    \algrule[\csname ALG@ind@\the\ALG@printindent@tempcnta\endcsname]%
    \advance \ALG@printindent@tempcnta 1
    \ifnum \ALG@printindent@tempcnta<\numexpr\theALG@nested+1\relax
    \repeat
    \fi
    \fi
}
\patchcmd{\ALG@doentity}{\noindent\hskip\ALG@tlm}{\ALG@printindent}{}{\errmessage{failed to patch}}
\patchcmd{\ALG@doentity}{\item[]\nointerlineskip}{}{}{} 
\makeatother

\allowdisplaybreaks

\begin{document}

\title{Harnessing Chaotic Signals for\\ Wireless Information and Power Transfer}

\author{Authors}
\author{Priyadarshi Mukherjee, \textit{Senior Member, IEEE}, Constantinos Psomas, \textit{Senior Member, IEEE},\\ and Ioannis Krikidis, \textit{Fellow, IEEE}
\thanks{P. Mukherjee is with the Department of Electrical Engineering and Computer Science, Indian Institute of Science Education and Research  Bhopal, India (e-mail: priyadarshi@ieee.org).

C. Psomas is with the Department of Computer Science and Engineering, European University  Cyprus, Cyprus (e-mail: c.psomas@euc.ac.cy).

I. Krikidis is with the Department of Electrical and Computer Engineering, University of Cyprus, Cyprus (e-mail: krikidis@ucy.ac.cy).}}

\maketitle

\begin{abstract}
Chaotic dynamical systems have attracted considerable attention due to their inherent randomness and high sensitivity to initial conditions, which makes them ideal for secure wireless communications. Beyond security, these same characteristics also make chaotic signals particularly effective for wireless power transfer (WPT) applications. On the other hand, connectivity along with self-sustainability are the two cornerstones of the upcoming sixth generation (6G) standard for radio communications. Consequently, with the massive increase in wireless devices and sensors, the concept of self-sustainable wireless networks is becoming more relevant. The aspect of WPT to the widely spread wireless devices and simultaneous wireless information and power transfer (SWIPT) among these devices will play a crucial role in the 6G communication systems. In this context, it has been experimentally observed that chaotic signals result in better WPT performance as compared to the existing benchmark schemes. Hence, in this paper, we characterize the generalized WPT performance of the multi-dimensional chaotic signals and present the use case of the Lorenz and the H\'enon chaotic systems. Moreover, we provide a novel differential chaos shift keying (DCSK)-based WPT receiver architecture ideal for enhanced energy harvesting (EH). Furthermore, we propose DCSK-based transmit waveform designs for multi-antenna SWIPT architectures and investigate the impact of the bit error rate-harvested energy trade-off. Our goal is to explore these aspects of the chaotic signals and discuss their relevance in the context of both WPT and SWIPT.
\end{abstract}

\section{Introduction}

Wireless traffic is growing at an explosive rate in recent years and it is expected to increase more than five times between $2023$ and $2028$ \cite{ericsson}. Applications such as energy-sustainable Internet of Things (IoT) networks and massive machine-type communications are becoming increasingly relevant in today's beyond fifth generation (B5G) wireless networks and as well as in upcoming sixth-generation (6G) communication systems. In such scenarios, a large number of devices are spread over a geographical region, and so the aspect of charging or powering them is a very critical issue. Moreover, the limited energy resources of these devices and the requirement for self-sustainability play a very important role in determining the frequency of their recharging. Furthermore, the fact that they are often placed in remote areas like forests or ``dead'' zones in mountainous regions makes them susceptible to degraded quality-of-service (QoS) performance. Consequently, the usage of renewable energy sources such as solar or wind energy is appearing to be a viable option. However, due to the costly, unfeasible in some scenarios, and uncontrollable nature of these alternatives, the possibility of considering wireless power transfer (WPT) as an option is gaining more and more traction. As a result, WPT-enabled 6G networks are considered a key component of upcoming energy-sustainable communication systems and several standardization activities have already been initiated. For instance, the 3GPP “Service and System Aspects” working group has started discussions on ambient energy harvesting (EH) IoT devices \cite{stndrd}. Here, the devices are powered wirelessly by harvesting energy from ambient and/or designated radio frequency (RF) signals \cite{newref}. This is achieved by using a rectenna, i.e., rectifying antenna at the receiver, which converts the received radio-frequency (RF) signals to direct current (DC). In this article, we explore the possibility of harnessing chaotic signals, a special class of signals, for wirelessly charging devices. We demonstrate how chaotic signals can be effectively exploited for the purpose of enhanced WPT. Moreover, we show how next-generation chaotic receivers can be designed for sustainable 6G networks and integrated without compromising WPT efficiency or information-transfer performance.

\begin{figure*}
    \centering
    \includegraphics[width=0.68\linewidth]{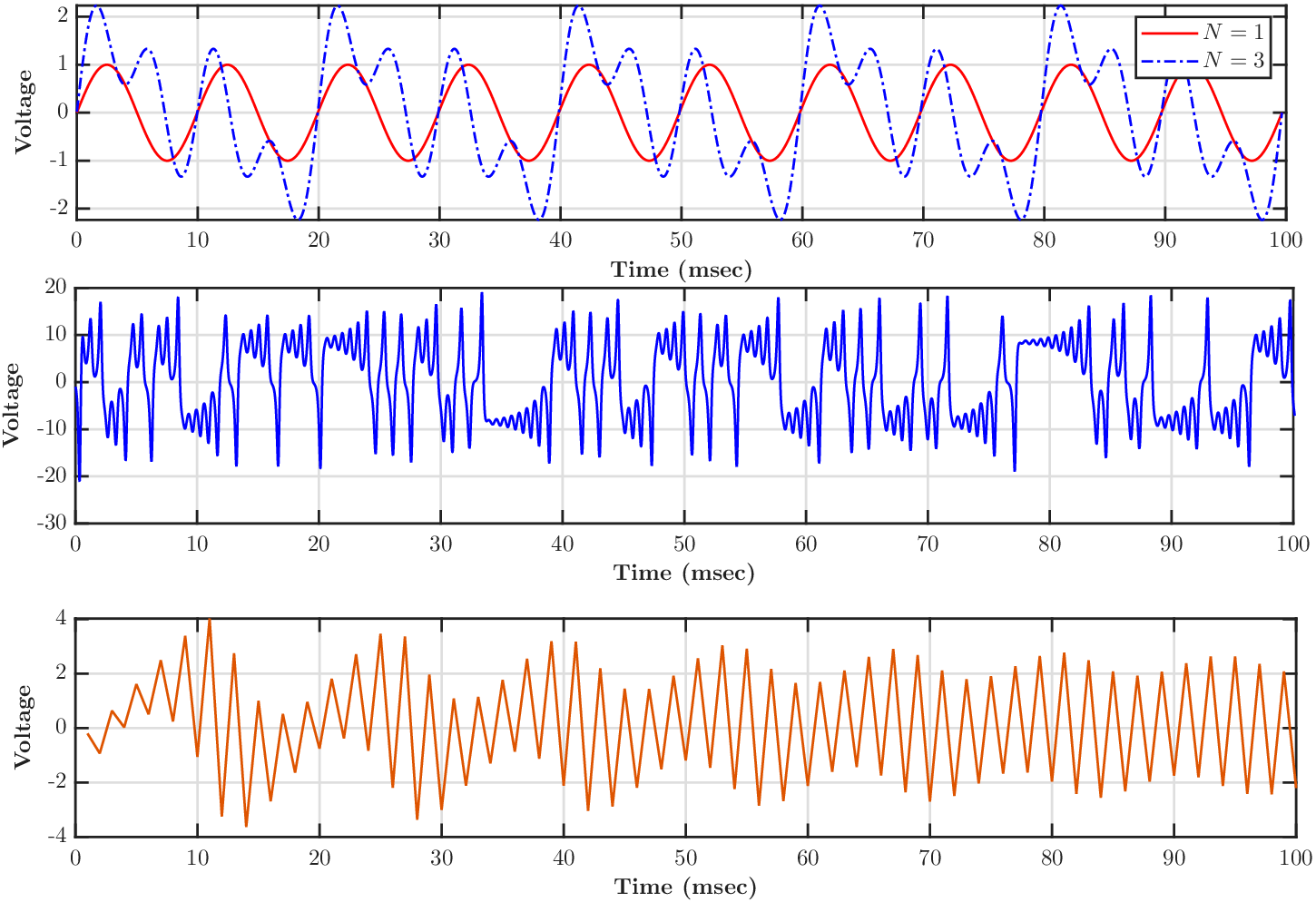}
    \caption{(a) $N$-tone sinusoid, (b) Lorenz signal, and (c) H\'enon signal.}
    \label{suma}
\end{figure*}

Note that, there exists a significant volume of research in the literature, which aims at the maximization of a system's end-to-end WPT efficiency. This, in turn, depends on  factors such as exploiting the nonlinear characteristics of the EH circuits at the receiver. Specifically, a typical EH circuit consists of a diode followed by a low pass filter. Therefore, the reason behind the nonlinearity of the EH circuit is attributed to the behavior of the diode. However, it is important to observe that the choice of appropriate transmit waveforms and their associated parameters plays a pivotal role in characterizing the WPT performance of the system. The authors in \cite{chaosexp2} experimentally demonstrate that, even with a constant transmit power, proper selection of transmit waveforms is extremely important for enhancing the WPT performance of the EH circuit. Specifically, the nonlinearity of the EH circuit causes waveforms with high peak-to-average-power-ratio (PAPR), like the orthogonal frequency division multiplexing (OFDM)-based signal and the chaotic signal, result in enhanced DC output. However, it is important to remember that signals with high PAPR are not desired in conventional communication systems. The reason for this being the fact that they suffer distortions
when amplified at the power amplifier of the transmitter.

Although the work in \cite{chaosexp2} experimentally demonstrates the benefits of chaotic signals, no proper analytical framework exists to support this observation. Moreover, the investigation of a generalized theoretical perspective is required to evaluate and characterize the process of energy transfer for chaotic systems. Therefore, based on the proposed framework, we evaluate the performance of chaotic signal-based WPT for the use cases of the Lorenz and the H\'enon systems. Furthermore, we investigate a chaotic modulation scheme, namely, the differential chaos shift keying (DCSK)-based WPT and propose a new receiver architecture, which facilitates the EH process. Specifically, we propose DCSK-based novel transmit waveform designs, which facilitate the energy transfer performance. Finally, by considering a DCSK-based wireless transmitter, we present two new low-complexity receiver architectures for simultaneous wireless information and power transfer (SWIPT). In the context of these receiver architectures, we propose novel DCSK-based transmit waveform designs, which take into account the wireless channel statistics and the system parameters. By doing so, we take care of the stringent QoS requirements of both information and energy transfer.

The rest of the article is organized as follows. First, we discuss the generalized chaotic signal and DCSK-based framework for WPT, where we also propose a novel receiver architecture. We then discuss DCSK-based transmit waveform designs for SWIPT, focusing on a single-input multiple-output (SIMO) architecture first, and a self-sustainable reconfigurable intelligent surface (RIS)-aided topology second. We conclude with an extensive discussion on opportunities and future directions related to the fascinating and under-explored field of chaotic signals and its applications in the upcoming 6G communication systems.

\section{Chaotic Waveforms for WPT}

In this section, we consider a simple point-to-point communication scenario, where the transmitter is a chaotic signal generator and the receiver consists of an EH circuit that converts the received signal to DC. We discuss about generalized chaotic system-based WPT, the practical limitations of such systems, and finally, we propose a DCSK-based receiver design along with its associated transmit waveform designs.

\subsection{Generalized Chaotic WPT}

A chaotic system is mathematically represented as a set of coupled differential equations. In other words, it is a  multi-variable system, where the variables influence each other, causing their temporal rates of change to depend on each other's values. This implies that the choice of initial state is very crucial and hence, cannot be overlooked. Also, we cannot independently solve a single equation of chaotic systems and need to consider the entire system to find the solutions for all variables. There are numerous such systems available in the literature but in this work, we consider the continuous time Lorenz system and the discrete time H\'enon system as use cases and investigate their WPT performances. The conventional Lorenz chaotic system is represented as follows  \cite{wcl2}
\begin{align} \label{ldef}
\dot{x}&=\sigma(y-x), \nonumber \\
\dot{y}&=x(r-z)-y, \nonumber \\
\dot{z}&=xy-\beta z, 
\end{align}
where $x,y,z$ are state variables, $\sigma,r,\beta>0$ are control parameters, and $\dot{x},\dot{y},$ and $\dot{z}$ denote $\frac{dx}{dt},\frac{dy}{dt},$ and $\frac{dz
}{dt}$, respectively. A conventional H\'enon system is expressed as \cite{wcl2}
\begin{align} 
x_{n+1}&=y_n+1-\gamma x_n^2, \nonumber\\
y_{n+1}&=\delta x_n,
\end{align}
where $x,y$ are state variables and $\gamma,\delta$ are control parameters. For a certain set of system parameters, Fig. \ref{suma} demonstrates the uniqueness of chaotic signals and how they are different from the conventional multisine signals. Specifically, for the Lorenz system, we plot the variation of $x$ with $\sigma=10,b=\frac{8}{3},r=28$ and initial point $x_0=-1,y_0=-2,z_0=1.5$ in Fig. 1b. Similarly, for the H\'enon system, we demonstrate the variation of $x$ with $\gamma=0.96,\delta=0.2$ and initial point $x_0=-0.2,y_0=-2$ in Fig. 1c. Both the figures clearly justify the `chaotic' nature of the signals and also, it is obvious from the figures that, in either cases, a different initial point will lead to a different trajectory.

\begin{figure}[!t]
\centering\includegraphics[width=\linewidth]{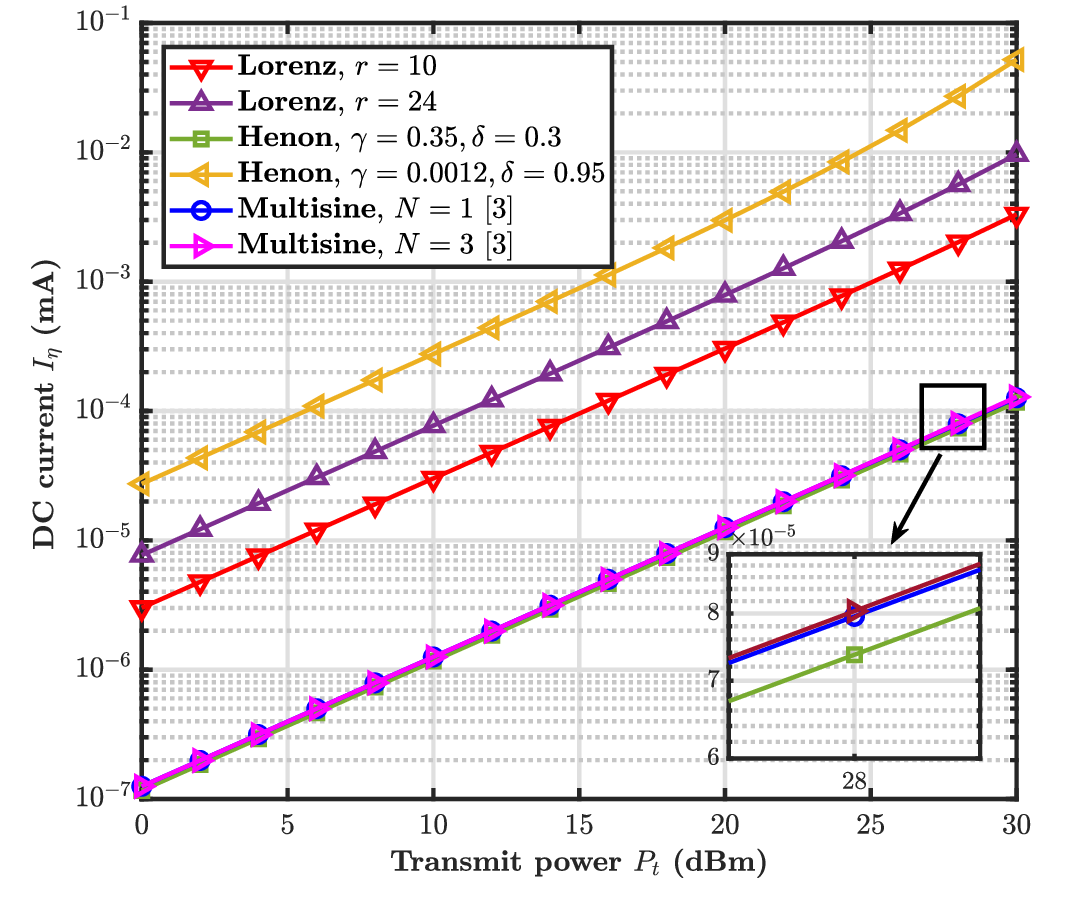}
\caption{Performance comparison of the chaotic waveforms (Lorenz, Henon) with conventional multisine signals.
}
\label{fig:mult}
\end{figure}

For both the considered systems, we assume $x$ as the transmitted signal and thereafter, by considering the nonlinear EH model at the receiver, we obtain the output DC current as $I_{\eta}$. For obtaining insights on the WPT performance of any chaotic system, we first obtain its equilibrium points and then, investigate the system's behavior around them \cite{wcl2}. Thereafter, the harvested DC is evaluated in terms of the received signal and the associated system parameters; for example, $\sigma,r,$ and $\beta$ for the  Lorenz system and $\gamma$ and $\delta$ for the H\'enon system, respectively. Fig. \ref{fig:mult} demonstrates the WPT performance of both Lorenz and H\'enon chaotic systems and also compare them with the existing $N$-tone multisine waveforms. We observe that, for the Lorenz system, the harvested energy increases with increasing $r$. However, note that, $r$ cannot be indefinitely increased as it depends on the stability of the system \cite{wcl2}. The figure also shows that the choice of parameters is equally important as the choice of waveforms; notice the performance gap for the H\'enon map between $\left(\gamma=0.35,\delta=0.3\right)$ and $\left(\gamma=0.0012,\delta=0.95\right)$. These considered values of $\gamma$ and $\delta$ have been considered by performing a stability analysis and observing the system behavior in the corresponding stable region of operation. Hence, we conclude from the figure that random choice of waveform from the chaotic family with an arbitrary set of system parameters does not necessarily guarantee an enhanced WPT performance.

However, the key problem for the family of chaotic systems discussed above is the requirement of perfect transmitter-receiver pair synchronization. Moreover, a crucial challenge with respect to embedded designing of such systems is achieving an acceptable level of precision with practical hardware constraints such as memory size and computational capacity \cite{cdisc}. To address these issues, a number of digital chaotic modulation schemes have been proposed in the literature. In this context, we next look at the noncoherent simple-to-implement DCSK modulation technique.

\subsection{DCSK-based WPT}  \label{cdsk}

\begin{figure}[!t]
\centering\includegraphics[width=\linewidth]{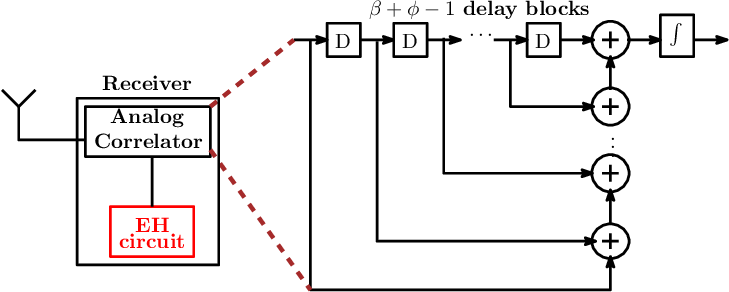}
\caption{Proposed receiver architecture.}
\label{jstsp1}
\end{figure}

Specifically, a variant of DCSK, namely, the short-reference DCSK (SR-DCSK) is proposed in \cite{srdcsk}. If $x_{l,k}$ is the chaotic sequence of length $\phi$, then during the $l$-th transmission interval, a SR-DCSK symbol of length $\beta+\phi$ is represented as \cite{srdcsk}
\begin{align}  \label{symsr}
& s_{l,k} \nonumber\\
 &=\!\!\begin{cases} 
x_{l,k}, & \!\!\! k=(l-1)(\beta+\phi)+1,\dots,(l-1)\beta+l\phi,\\
d_lx_{l,k-\phi}, & \!\!\!k=(l-1)\beta+l\phi+1,\dots,l(\beta+\phi),
\end{cases}&
\end{align}
where $d_l=\pm1$ is the information symbol. In the literature, $\beta$ is defined as the \textit{spreading factor} and the special case of $\beta=\phi$ leads to the conventional DCSK symbol construction. Therefore, DCSK requires generating a long reference signal equal in duration to the data signal. On the contrary, SR-DCSK uses a much shorter reference sequence (e.g., fractional duration), which saves symbol time. This reduction in sequence length minimizes processing delays and achieves higher data rates and that is the reason why we have considered SR-DCSK in this work. Although various chaotic maps exist for the purpose of $x_{l,k}$ generation, we consider the Chebyshev map for its excellent ``noise-like'' correlation properties \cite{cheby}. Notably, in this context, it implies that the Chebyshev map generated chaotic signals with different initial values to be regarded as quasi-orthogonal.

By considering SR-DCSK, we propose various transmit waveform designs and WPT-based receiver architectures, which enhance the EH performance of the end-to-end communication system. Moreover, as demonstrated in Fig. \ref{jstsp1}, we propose placing a $\beta+\phi$ bit analog correlator at the receiver, prior to the EH unit \cite{jstsp}. An ideal $\left( \beta+\phi \right)$-bit analog correlator consists of $\beta+\phi-1$ number of delay blocks, which results in minimal hardware complexity \cite{anaco2}. In other words, we require a $2\beta$ bit correlator for the conventional DCSK-based scenario. Note that, an analog correlator is a low complexity yet very effective signal integrator, which simply sums the received signal over a certain time interval. Specifically, this is different from a conventional summing circuit; a summing circuit generates an output, which is the sum of multiple input signals whereas the analog correlator allows an effective integration of the same signal over a certain time interval, i.e. it generates an output, which is the sum of the delayed versions of the same signal. Moreover, the analog correlator is essentially an active device, but with a nominal power consumption. Specifically, it is observed that a CMOS-implemented 11-bit analog correlator consumes about 40 nW of power \cite{anaco2} Hence, we have not considered this minimal energy consumption in the context of applying the correlator. Furthermore, considering an analog correlator over its digital counterpart eliminates the requirement of any clock generator as well as any synchronization issues. The noise contribution from the delay cells is kept significantly smaller as compared to the front-end noise and this has been practically implemented in \cite{anaco2}. Thereafter, it is important to note that the resulting noise power is not a matter of concern here as it results in negligible harvested energy at the receiver. The motivation behind the employment of this block is two-fold: i) it enables controlling the signal at the input of the harvester, and ii) it enhances the PAPR of the signal \cite{jstsp}. Accordingly, for a given $\beta$, we make a proper selection of $\phi$, which results in an enhanced WPT performance.

\begin{figure}[!t]
\centering\includegraphics[width=\linewidth]{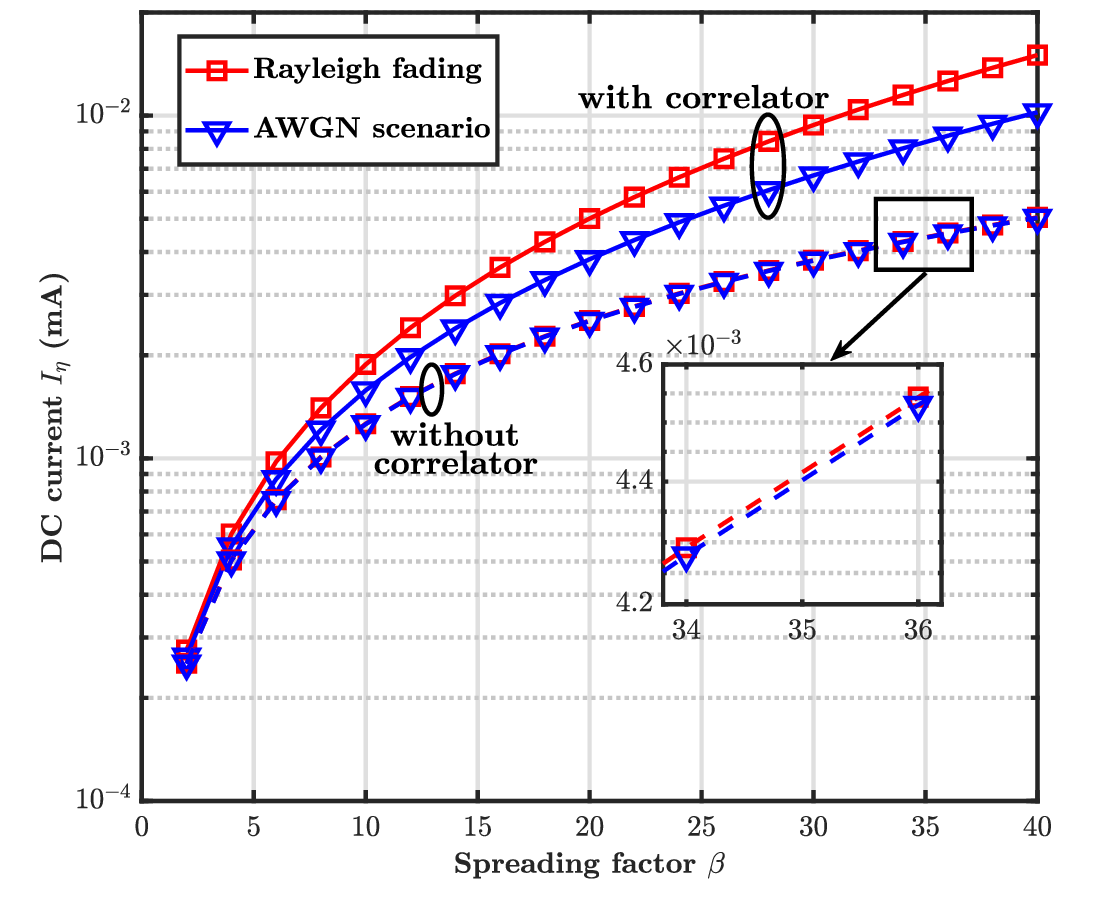}
\caption{Impact of proposed receiver architecture.}
\label{jstsp2}
\end{figure}

Fig. \ref{jstsp2} illustrates the impact of the proposed DCSK-based WPT architecture in a generalized flat fading scenario. We observe significant performance gains achieved due to the presence of the analog correlator at the receiver. The figure also shows that for a given $\beta$, the EH performance deteriorates as the wireless channel moves from being Rayleigh faded to an additive white Gaussian noise (AWGN)-based scenario. Therefore, unlike in the information transfer scenario, fading is beneficial for EH. Further analytical investigation demonstrates the importance of transmit waveform design in the context of our proposed receiver architecture. Specifically, for a given $\beta$, we vary $\phi$ such that $\frac{\beta}{\phi} \in \mathbb{Z}^+$ and show its impact on the energy transfer performance. We observe that the maximum $I_{\eta}$ is obtained at $\phi=1$ followed by a monotonically deteriorating performance. The reason for this observation is attributed to the fact that with increasing $\phi$, the degree of \textit{randomness} among the consecutive chaotic components decreases and vice-versa. Finally, for the special case of $\phi=\beta$, the WPT performance coincides with that of the conventional DCSK, where the second half of the symbol is the identical/inverted replica of the first half.

\section{Chaotic Waveforms for SWIPT}

Since chaotic waveforms are beneficial for WPT, it is interesting to investigate the aspect of chaotic signal-based waveform designs for SWIPT. Note that, energy and information transfer are two conflicting tasks and as a result, a single transmit waveform  cannot be simultaneously optimal for both WPT and information transfer (IT). Henceforth, we propose two DCSK-based SWIPT architectures and investigate the aspect of appropriate transmit waveform designs in a generalized frequency selective fading scenario. Unless otherwise stated, we consider a frequency selective scenario with two independent paths and corresponding power gains $\Omega_1,\Omega_2$ such that $\Omega_1+\Omega_2=1.$

\subsection{Waveform Design for DCSK-based SWIPT}

We consider a single-input multiple-output (SIMO)  SWIPT set-up, with a single
antenna transmitter and an $N$ antenna receiver. The transmitter employs a DCSK-based signal generator and the receiver is a multi-antenna architecture, where each antenna switches between IT or EH modes depending on the desired requirements \cite{twc}. Suppose $M(K)$ among the $N$ receiver antennas are selected for EH (IT) such that $M+K=N$. During each chaotic symbol transmission, all the $K$ antennas are individually connected to a conventional SR-DCSK-based demodulator \cite{srdcsk}. The output from all these $K$ demodulators are combined and compared with a threshold to recover the actual transmitted data. The output from the remaining $M$ antennas are fed to a parallel-in serial-out shift register, which in turn acts as the input to the analog correlator followed by the EH unit. Depending on the application specific requirements, we look into the transmit waveform designs for a given system configuration; for a given $\beta$, we intend to find out the acceptable $\phi$.

Moreover, we investigate an interesting trade-off in this work. We know that the performance of conventional SWIPT systems is characterized by investigating the fundamental rate-energy trade-off. This is an effective way to quantify the performance of conventional SWIPT systems and compare its various architectures. On the contrary, chaotic waveform-based codewords are finite length in nature and thus, investigating the conventional rate-energy trade-off is not enough for complete characterization of chaotic SWIPT systems. As a result, we investigate the bit error rate (BER)-energy tradeoff for such systems and also its impact on the transmit waveform design. Note that this fundamental tradeoff can be studied from the information theory perspective, but here we adopt a more practical metric, which is the BER-energy tradeoff. Specifically, we characterize the the receiver performance in terms of the achievable success rate (SR) - harvested energy region, where we define ${\rm SR}=1-{\rm BER}$.

\begin{figure}[!t]
\centering\includegraphics[width=\linewidth]{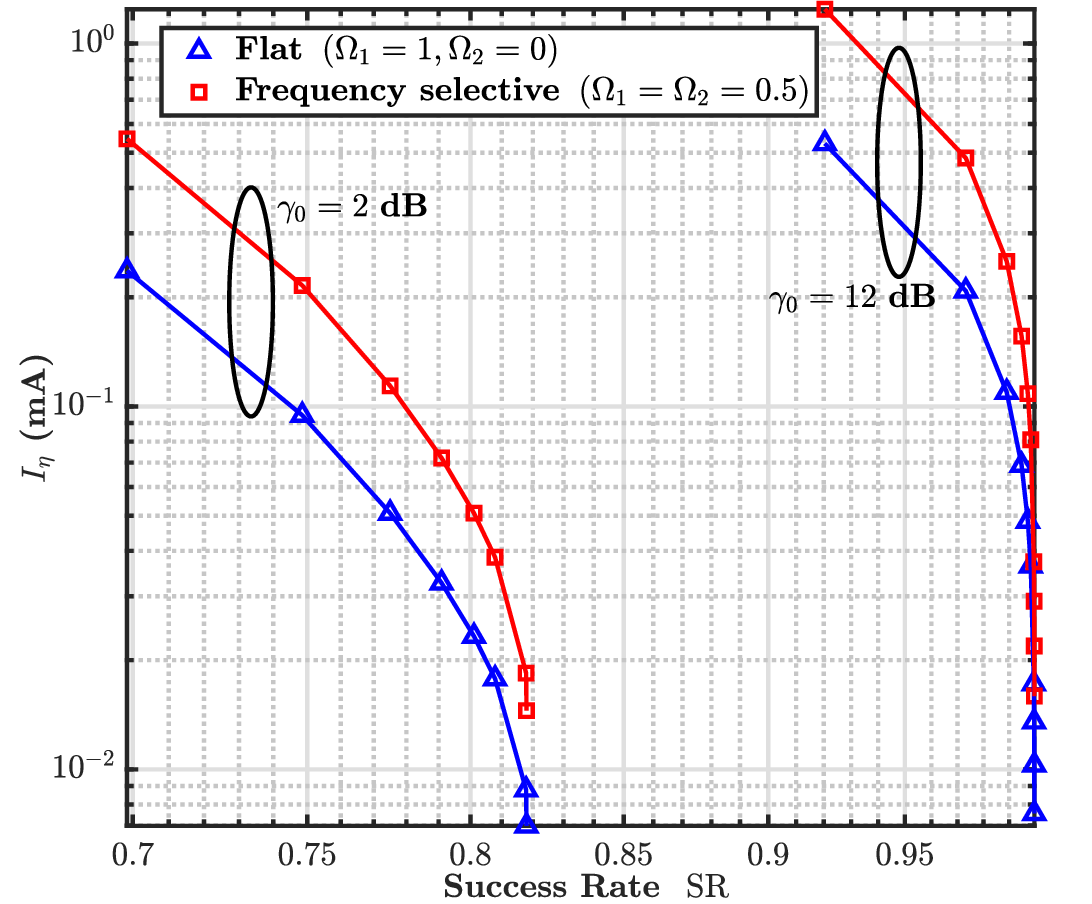}
\caption{Characterization of the ${\rm SR}-I_{\eta}$ trade-off.}
\label{tcom2}
\end{figure}

Fig. \ref{tcom2} demonstrates the ${\rm SR}-I_{\eta}$ trade-off with the system parameters $N=5,M=1,K=4,$ and $\beta=60$ in a Rayleigh fading environment. We consider two different path-loss dependent signal-to-noise ratio at the receiver, specifically, $\gamma_0=2,10$ dB for the proposed SWIPT architecture. Here, two different scenarios have been taken into account: flat fading and frequency selective fading. We observe that, for any considered scenario, increasing $\phi$ implies enhancement of the BER performance, while the EH performance degrades. The figure illustrates that a frequency selective scenario is better for energy transfer as compared to its flat fading counterpart; with all the other parameters remaining constant, observe the performance gap between the flat fading scenario $\left( \Omega_1=1,\Omega_2=0 \right)$ and the frequency selective scenario $\left( \Omega_1=\Omega_2=0.5\right)$. The figure also shows that a lower $\gamma_0$ degrades both the energy and information transfer performance at the receiver. Hence, we conclude that the choice of $\phi$ cannot be arbitrarily done and it jointly depends on the application-specific requirements, parameters of the transmit waveform, the channel statistics, and also on the number of receiver antennas being utilized in the EH and IT mode, respectively.

\subsection{Waveform Design for Self-Sustainable RIS-Aided DCSK-based SWIPT}


Here, we consider a scenario with no direct link between the transmitter and the receiver due to heavy obstruction from obstacles. As a result, to realize the communication process, a self-sustainable RIS \cite{zeris} is employed. In this RIS, the reflecting elements (REs) are grouped into two parts: EH and IT. Specifically, the beam-steering functionality of the IT portion requires a sufficient amount of energy consumption, which in turn, is provided by the energy harvested by the EH portion. Therefore, if the chaotic signal-based waveforms are used for communication, it will lead to better energy transfer performance at the RIS. Hence, lesser number of REs can constitute the EH part of the RIS and a larger fraction of the RIS can be used for the communication purpose. For this purpose, we propose an $N$ element self-sustainable RIS-aided single-input single-output (SISO) setup, where the transmitter employs a SR-DCSK-based signal generator \cite{tcom} and the receiver is the conventional SR-DCSK-based demodulator. There are $M$ and $K$ REs in the EH and IT section of the RIS, respectively, such that $M+K=N$. Moreover, we propose a combining approach in the EH section of the RIS, where each individual RE has its own EH unit. Furthermore, as discussed earlier in Section \ref{cdsk}, we also employ an analog correlator prior to the EH unit attached to each RE. Identical to the previously discussed scenario, we investigate the importance of DCSK-based transmit waveform designs. The objective of our study is to guarantee that the energy harvested at the EH section must be sufficient to meet the energy requirements of the IT section, such that it is successfully able to provide a satisfactory BER performance.

However, a self-sustainable RIS suffers from two disadvantages. Firstly, both the transmitter and the receiver need to be on the same side of the RIS and secondly, the reflected signal suffers from double attenuation, which severely degrades the quality of the signal reaching the receiver. By taking these factors into account, the work in \cite{mfris1} proposed a generalized multi-functional (MF) RIS. An MF-RIS harvests energy from the incident signal to enable multidimensional and full-space signal manipulation. Specifically, the harvested energy is used for simultaneous reflection and transmission of the amplified incident signals towards the desired users. In this context, the work in \cite{wcl3} investigates the aspect of transmit waveform design in a DCSK-based MF-RIS-aided framework. This is both important and interesting, because as stated earlier, chaotic signal-based waveform imply better WPT performance. Therefore, a larger fraction of the MF-RIS can be utilized by the transmitter for communicating with the users (with an acceptable BER performance) present on both the sides of the MF-RIS.

\section{Opportunities and Future Directions}
In this era of B5G and upcoming 6G networks, connectivity with self-sustainability is the key. As a result, zero-energy communication devices are being looked up to as the building blocks of future wireless networks. In this context, the family of chaotic signals opens up an under-explored research direction of chaotic signal-based next generation wireless communication networks. From the above discussion, it is clear that a narrow focus on chaotic signals and the appropriate choice of its system parameters results in a significant performance enhancement in both WPT and SWIPT. Therefore, from fundamentals to applications, we identify the following potential future directions.

\subsection*{$\circ$ Mathematical Framework for Chaotic Systems}

Existing mathematical models do not provide a generalized framework for solving chaotic systems. Some approaches like variational iteration and homotopy perturbation methods exist for obtaining approximate analytical solution of a few members of this chaotic family, for example, the Lorenz system. In some cases, numerical techniques such as the higher order Runge–Kutta methodology provide an acceptable approximation for solving the chaotic systems. However, the investigation of sophisticated mathematical models, which accurately solve any chaotic system and also provide a balance between the practical hardware constraints such as memory size and computational capacity is still an open problem. A framework for solving any arbitrary chaotic system, if obtained, will be especially useful in the context of its application in the 6G self-sustainable zero-energy wireless networks.

\subsection*{$\circ$ Chaotic-based Modulation Schemes}

Current works that exploit chaotic signals for designing transmit waveforms, for both WPT and SWIPT, are primarily DCSK-based in nature. The popularity of DCSK lies in its non-coherent characteristics but in this case, the sole problem lies in its symbol structure. Specifically, only one half of the symbol duration is used for the transmission of information-bearing signals and the other half is basically a wasted resource. Therefore, the aspect of looking into new chaotic modulation schemes, which also leads to significant enhancement in both WPT and SWIPT performance should be explored. It is interesting to note that even for a fixed symbol length, the very design of the symbol affects the performance of the transmit waveform and depending on the application-specific requirements, this critical aspect must be considered.

\subsection*{$\circ$ Application to Emerging Technologies}

Equipped with new modulation schemes, the aspect of investigating chaotic signal-based communication strategies at the crossroads of SWIPT and other emerging technologies (e.g., integrated communication and sensing, fluid antenna systems, etc) is also open for future works. What makes these problems more interesting is the fact that as far as waveform designing is concerned, information and energy transfer are two contrasting criteria. A single particular waveform cannot be optimal for meeting both these objectives. It depends on the application-specific requirements and the criterion for self-sustainability that jointly decides the parameters of the transmit waveforms.

\subsection*{$\circ$ Design of Chaotic Receiver Architectures}

Improved and integrated WPT and SWIPT receiver architectures have been proposed in the recent years. However, this is not the case for their chaotic signal-based counterparts. Since the shape of the basis functions is not known beforehand, the conventional matched filter-based approach cannot be used for DCSK-based communication scenarios. Moreover, attaining perfect synchronization between the transmitter and the receiver circuits is extremely critical for generalized chaotic signal-based scenarios. As a result, circuit designing and implementation of chaotic signal-based WPT and SWIPT architectures remains an open problem for investigation. Furthermore, considering multiple antennas at the transmitter/receiver and the other practical constraints like noise, non-linearity, variations over process, voltage, and temperature can impact the correlator performance; this makes the scene even more challenging. For example, the noise contribution from the delay cells has to be kept low compared to the front-end noise, which accounts for most of the power consumption in the correlator. Finally, the practical difficulties of dealing with the output signal from the analog correlator, which has a significantly higher PAPR, is still an open as well as interesting research direction to pursue.

\subsection*{$\circ$ Standardization Efforts}

The class of ambient IoT networks comprises of wireless devices with ultra-low power budget. This power budget can be met by having dedicated EH units at the devices, which can harvest energy from the incoming RF signals. In this context, several standardization initiatives are underway; 3GPP and AirFuel Alliance have already initiated discussions and proposals on ambient IoT networks. Accordingly, integrating new chaotic signal–based waveforms, techniques, and receiver designs into these existing standards is a promising open research direction.

\section{Conclusion}

This article characterizes the usage of chaotic signals for both WPT and SWIPT. A generalized multi-dimensional chaotic signals-aided framework for WPT is provided, where the use cases of Lorenz and H\'enon chaotic systems have been considered. It has been shown that the selection of system parameters is equally important as the choice of signals. To improve the WPT performance of a DCSK-based system, a novel receiver architecture is proposed. It has been demonstrated that significant performance gain can be obtained by employing an analog correlator prior to the EH unit in the receiver. Although no detailed analysis is provided in this article, various characteristics of DCSK have been exploited to propose new transmit waveform designs in the context of various multi-antenna chaotic SWIPT architectures. These proposed waveform designs take into account the bit error rate-harvested energy trade-off such that a satisfactory BER as well as EH performance can be guaranteed. Finally, a list of opportunities and open directions are discussed as to how chaotic signals can play a pivotal role in the context of next generation zero-energy self-sustainable communication devices and the wireless communication in between. Overall, we foresee that chaotic signals will play an enabling role for the upcoming B5G and 6G networks, with the potential of making zero-energy wireless communication networks a reality.

\bibliographystyle{IEEEtran}
\bibliography{ref}
\end{document}